\documentstyle[prd,aps]{revtex}
\begin{document}
\input epsf
\draft
\newcommand{\picdir}[1]{./#1}     
\newcommand{\gpicdir}[1]{/home/felder/mypapers/hybrid/figs/#1}
\def\pbg{\hbox{${\mathbf \pi\beta\gamma}$}}
\def\tsigma{\tilde{\sigma}}
\def\lesssim{\mathrel{\hbox{\rlap{\hbox{\lower4pt\hbox{$\sim$}}}\hbox{$<$}}}}
\def\gtrsim{\mathrel{\hbox{\rlap{\hbox{\lower4pt\hbox{$\sim$}}}\hbox{$>$}}}}
 \twocolumn[\hsize\textwidth\columnwidth\hsize\csname
 @twocolumnfalse\endcsname
\preprint{CITA-2000-??, SU-ITP-00-??, CERN-TH/2000-??,  hep-ph/0012142}
\date{\today}
\title{\Large\bf Dynamics of Symmetry Breaking and Tachyonic Preheating}
\author{Gary Felder$^{1}$, ~Juan Garc\'\i a-Bellido$^{3}$, ~Patrick
B. Greene$^{2,4}$, ~Lev Kofman$^{2}$,\\ ~Andrei Linde$^{1}$ ~and ~Igor
Tkachev$^{5}$} \address{${}^1$Department of Physics, Stanford
University, Stanford, CA 94305, USA} \address{${}^2$ CITA, University of
Toronto, 60 St. George Street, Toronto, ON M5S 1A7, Canada}
\address{${}^3$Dep. F\'\i sica Te\'orica C-XI, Universidad Aut\'onoma
Madrid, Cantoblanco, 28049 Madrid, Spain} \address{${}^4$Dept. of
Physics, University of Toronto, 60 St. George Street, Toronto, ON M5S
1A7, Canada} \address{${}^5$Theory Division, CERN, CH-1211 Geneva 23,
Switzerland} \maketitle
\begin{abstract}
We reconsider the old problem of the dynamics of spontaneous
symmetry breaking using 3d lattice simulations, and develop a
theory of tachyonic preheating, which occurs due to the spinodal
instability of the scalar field. Tachyonic preheating is so
efficient that symmetry breaking typically completes within a
single oscillation of the field distribution as it rolls towards
the minimum of its effective potential. As an application of this
theory we consider preheating in the hybrid inflation scenario,
including SUSY-motivated F-term and D-term inflationary models. We
show that preheating in hybrid inflation is typically tachyonic
and the stage of oscillations of a homogeneous component of the
scalar fields driving inflation ends after a single oscillation.
Our results may also be relevant for the theory of the formation
of disoriented chiral condensates in heavy ion collisions.
\end{abstract}
\pacs{PACS: 98.80.Cq, CITA-2000-60, SU-ITP-00-35, IFT- UAM/CSIC-00-40,
FT-UAM-00/26, CERN-TH/2000-365  hep-ph/0012142} \vskip2pc]

\section{Introduction}

Spontaneous symmetry breaking is a basic feature of all
realistic theories of elementary particles. In the simplest
models, this instability appears because of the presence of
tachyonic mass terms such as $-m^2\phi^2/2$ in the effective
potential. As a result, long wavelength quantum fluctuations
$\phi_k$ of the field $\phi$ with momenta $k < m$ grow
exponentially, $\phi_k \sim \exp (t\sqrt{m^2- k^2})$, which leads
to spontaneous symmetry breaking.

This process may occur gradually, as in the theory of second order
phase transitions, when the parameter $m^2$ slowly changes from
positive to negative and the degree of symmetry breaking gradually
increases in time \cite{KL72}. Sometimes the symmetry breaking
occurs discontinuously, due to a first order phase transition
\cite{KL76}. But there is also another possibility, which we will
study in this paper: The tachyonic mass term may appear suddenly,
on a time scale that is much shorter than the time required for
symmetry breaking to occur.  This may happen, for example, when
the hot plasma created by heavy ion collisions in the `little Big
Bang' suddenly cools down \cite{Bjorken}. A more important
application from the point of view of cosmology is the process of
preheating in the hybrid inflation scenario \cite{Hybrid,hybrid1},
where inflation ends in a `waterfall' regime triggered by
tachyonic instability.

During the last few years we have learned that particle production
by an oscillating scalar field may occur within a dozen
oscillations due to the nonperturbative process of particle
production called preheating~\cite{KLS}. Usually preheating is
associated with broad parametric resonance in the presence of a
coherently oscillating inflaton field \cite{KLS}, but other
mechanisms are also possible, see e.g. \cite{Prokopec,inst}.

In this paper we will concentrate on what we call {\it tachyonic
preheating}, which occurs due to tachyonic (spinodal)
instabilities in the field responsible for the symmetry breaking.
The process of symmetry breaking has been studied before by
advanced methods of perturbation theory, see e.g. \cite{Boyan} and
references therein. However, spontaneous symmetry breaking is a
strongly nonlinear and nonperturbative effect. It usually leads to
the production of particles with large occupation numbers
inversely proportional to the coupling constants.  As a result the
perturbative description, including the Hartree and $1/N$
approximations, has limited applicability. It does not properly
describe rescattering of created particles and other important
features such as production of topological defects.

For further  theoretical understanding of the issue one should go
beyond perturbation theory. Fortunately, during the last few years
new methods of lattice simulations have been developed. They are
based on the observation that quantum states of bose fields with
large occupation numbers can be interpreted as classical waves and
their dynamics can be fully analyzed by solving relativistic wave
equations on a lattice \cite{lattice,FT}. In our paper we will
further develop these methods, including effects of
renormalization, and  apply them to the investigation of
spontaneous symmetry breaking and tachyonic preheating. A
significant advantage of these methods as compared to other
lattice simulations of quantum processes is that the
semi-classical nature of the effects under investigation allows us
to have a clear visual picture of all the processes involved. That
is why this paper is accompanied by several computer generated
movies that show the development of symmetry breaking in various
models.

We will show that tachyonic preheating can be extremely efficient.
In many models it leads to the transfer of the initial potential
energy density $V(0)$ into the energy of scalar particles within a
single oscillation. Contrary to standard expectations, the first
stage of preheating in hybrid inflation is typically tachyonic,
which means that the stage of oscillations of a homogeneous
component of the scalar fields driving inflation either does not
exist at all or ends after a single oscillation. A detailed
description of our results will be given elsewhere \cite{GBFKLT}.

\section{Tachyonic Instability  and symmetry breaking}

Symmetry breaking occurs due to tachyonic instability and may be
accompanied by the formation of topological defects. Here we will
consider two toy models that are prototypes for many interesting
applications, including symmetry breaking in hybrid inflation.

\subsection{Quadratic potential}

The simplest model of spontaneous symmetry breaking is based on the
theory with the effective potential
\begin{equation}\label{aB1}
V (\phi) = {\lambda\over 4}(\phi^2-{  v}^2)^2 \equiv  {m^4\over
4\lambda} - {m^2\over
2}\phi^2 + {\lambda\over 4} \phi^4 \ ,
\end{equation}
where $\lambda \ll 1$. $V(\phi)$ has a minimum at $\phi = \pm v$
and a maximum at $\phi = 0$ with curvature $V'' = -m^2$.

The development of tachyonic instability in this model depends on
the initial conditions. We will assume that initially the symmetry
is completely restored so that the field $\phi$ does not have any
homogeneous component, i.e. $\langle \phi \rangle= 0$. But then
$\langle \phi \rangle$ remains zero at all later stages, and for
the investigation of spontaneous symmetry breaking one needs to
find the spatial distribution of the field $\phi(x,t)$.  To
avoid this complication, many authors assume that there is a small
but finite initial homogeneous background field $\phi(t)$, and
even smaller quantum fluctuations $\delta\phi(x,t)$ that grow on
top of it. This approximation may provide some interesting
information, but quite often it is inadequate. In particular, it
does not describe the creation of topological defects, which, as
we will see, is not a small nonperturbative correction but an
important part of the problem.

For definiteness, we suppose that in the symmetric phase $\phi=0$
there are usual quantum fluctuations of the massless field with
the mode functions  ${1 \over \sqrt{2 k}}e^{-ikt +i{\vec k \vec
x}}$ and then at $t = 0$ we `turn on' the term $-m^2\phi^2/2$
corresponding to the negative mass squared $-m^2$. The modes with
$k = |{\vec k}|  < m$ grow  exponentially so
 the dispersion of these fluctuations can be estimated as
\begin{equation}
\langle \delta\phi^2 \rangle
 = {1 \over 4\pi^2 } \int\limits_0^{m}  dk\, k \, e^{2t\sqrt{m^2- k^2}}   \ .
\label{aBB}
\end{equation}

To get a qualitative understanding of the process of spontaneous
symmetry breaking, instead of many growing waves with momenta $k <
m$ in (\ref{aBB}) let us consider first a single sinusoidal wave
$\delta\phi = \Delta(t) \cos kx$ with $k \sim m$ and with initial
amplitude $\sim {m\over 2\pi}$ in one-dimensional space. The
amplitude of this wave grows exponentially until its becomes
${\cal O}(v) \sim m/\sqrt \lambda$.  This leads to the division
of the universe into domains of size ${\cal O}(m^{-1})$ in which
the field changes from ${\cal O}(v)$ to ${\cal O}(-v)$. The
gradient energy density of domain walls separating areas with
positive and negative $\phi$ will be $\sim k^2\delta\phi^2=
O(m^4/\lambda)$. This energy is of the same order as the total
initial potential energy of the field $V(0) = m^4/4\lambda$. This
is one of the reasons why any approximation based on perturbation
theory and ignoring topological defect production cannot give a
correct description of the process of spontaneous symmetry
breaking.

Thus a substantial part of the false vacuum energy $V(0)$ is
transferred to the gradient energy of the field $\phi$ when it
rolls down to the minimum of $V(\phi)$. Because the initial state
contains many quantum fluctuations with different phases growing
at a different rate, the resulting field distribution is very
complicated, so it cannot give all of its gradient energy back and
return to its initial state $\phi = 0$.  This is one of the
reasons why spontaneous symmetry breaking and the main stage of
preheating in this model may occur within a single oscillation of
the field~$\phi$.

Meanwhile if one were to make the usual assumption that initially
there exists a small homogeneous background field $\phi \ll v$
with an amplitude greater than the amplitude of the growing
quantum fluctuations $\delta\phi$, so that $m/2\pi \ll \phi <
m/\sqrt \lambda$, one would find out that when $\phi$ falls to the
minimum of the effective potential the gradient energy of the
fluctuations remains relatively small.  One would thus come to
the standard conclusion that the field should experience many
fluctuations before it relaxes near the minimum of $V(\phi)$. To
avoid this error, we need to perform a complete study of the
growth of all tachyonic modes and their subsequent interaction
without making this simplifying assumption about the existence of
the homogeneous field $\phi$.

 Consider the tachyonic growth of all fluctuations with $k <
m$, i.e. those that contribute to $\langle \delta\phi^2 \rangle$
in Eq. (\ref{aBB}). This growth continues until $\sqrt{\langle
\delta\phi^2 \rangle}$ reaches the value $\sim v/2$, since at
$\phi \sim v/\sqrt 3$ the curvature of the effective potential
vanishes and instead of tachyonic growth one has the usual
oscillations of all the modes. This happens within the time $t_*
\sim {1\over 2 m} \ln{\pi^2\over \lambda}$. The exponential growth
of fluctuations up to that moment can be interpreted as the growth
of the occupation number of particles with $k \ll m$. These
occupation numbers at the time $t_*$ grow up to
\begin{equation}
n_k \sim \exp(2mt_*) \sim
\exp\left({\ln{\pi^2\over \lambda}}\right) = {\pi^2\over \lambda} \gg
1\,.
\end{equation}
One can easily verify that $t_*$ depends only logarithmically
on the choice of the initial distribution of quantum fluctuations. For small $\lambda$ the fluctuations with $k \ll m$ have very large occupation numbers,
and therefore they can be interpreted as classical waves of the
field $\phi$.

The dominant contribution to $\langle \delta\phi^2 \rangle$ in Eq. (\ref{aBB}) at the moment $t_*$ is given by the modes with wavelength $l_* \sim 2\pi
k_*^{-1} \sim \sqrt 2 \pi m^{-1} \ln^{1/2} {(C\pi^2/\lambda)} >
m^{-1}$, where $C = {\cal O}(1)$. As a result, at the moment when the
fluctuations of the field $\phi$ reach the minimum of the
effective potential, $\sqrt{\langle\phi^2\rangle} \sim v$, the
field distribution looks rather homogeneous on a scale $l \lesssim
l_*$. On average, one still has $\langle \phi\rangle = 0$. This
implies that the universe becomes divided into domains with two
different types of spontaneous symmetry breaking, $\phi \sim \pm
v$. The typical size of each domain is $l_*/2 \sim {\pi\over \sqrt
2 } \ m^{-1} \ln^{1/2}{C\pi^2\over \lambda}$, which differs only
logarithmically from our previous estimate $m^{-1}$. At later
stages the domains grow in size and percolate (eat each other up),
and spontaneous symmetry breaking becomes established on a
macroscopic scale.

Of course, these are just simple estimates which should be
followed by a detailed quantitative investigation. When the field
rolls down to the minimum of its effective potential, its
fluctuations scatter off each other as classical waves. It is
difficult to study this process analytically, but fortunately one
can do it numerically using the method of lattice simulations
developed in \cite{lattice,FT}. We performed our simulations on
lattices with either $128^3$ and $256^3$ gridpoints.  A detailed
description of our calculations will be given in \cite{GBFKLT};
here we will only present our main results.

\begin{figure}[Fig001]
\centering \leavevmode\epsfysize=8.5cm \epsfbox{\picdir{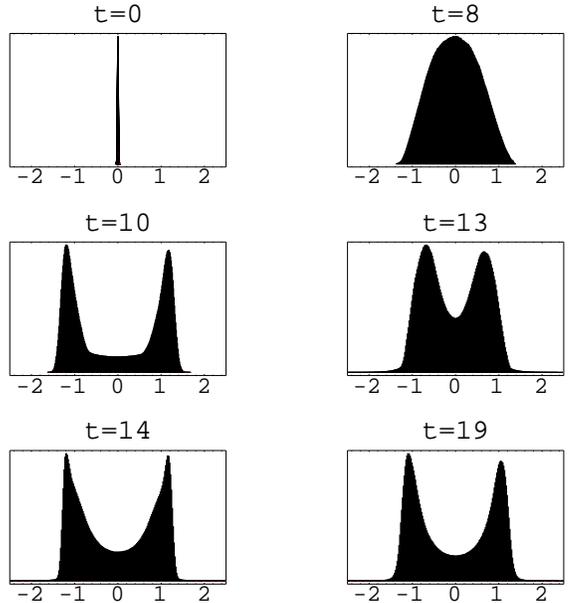}}

\

\caption[Fig001]{\label{onefielddistrib} The process of symmetry
breaking in the model (\ref{aB1}) for $\lambda = 10^{-4}$. In the
beginning the distribution is very narrow. Then it spreads out and shows
two maxima which oscillate about $\phi = \pm v$ with an amplitude much
smaller than $v$. These maxima never come close to the initial point
$\phi = 0$. The values of the field are shown in units of $v$.}
\end{figure}

Figure \ref{onefielddistrib} illustrates the dynamics of symmetry
breaking in the model (\ref{aB1}). It shows the probability
distribution $P(\phi,t)$, which is the fraction of the volume
containing the field $\phi$ at a time $t$ if at $t = 0$ one begins
with the probability distribution concentrated near $\phi = 0$,
with the quantum mechanical dispersion (\ref{aBB}).

As we see from this figure, after the first oscillation the
probability distribution $P(\phi,t)$ becomes narrowly concentrated
near the two minima of the effective potential corresponding to
$\phi = \pm v$. In this sense one can say that symmetry breaking
completes within one oscillation. To demonstrate that this is not
a strong coupling effect, we show the results for the model
(\ref{aB1}) with $\lambda = 10^{-4}$. Note that only when the
distribution stabilizes and the domains become large can one use
the standard language of perturbation theory describing scalar
particles as excitations on a (locally) homogeneous background.
That is why the use of the nonperturbative approach based on
lattice simulations was so important for our investigation.

\begin{figure}[Fig001]
\centering \leavevmode\epsfysize=7cm \epsfbox{\picdir{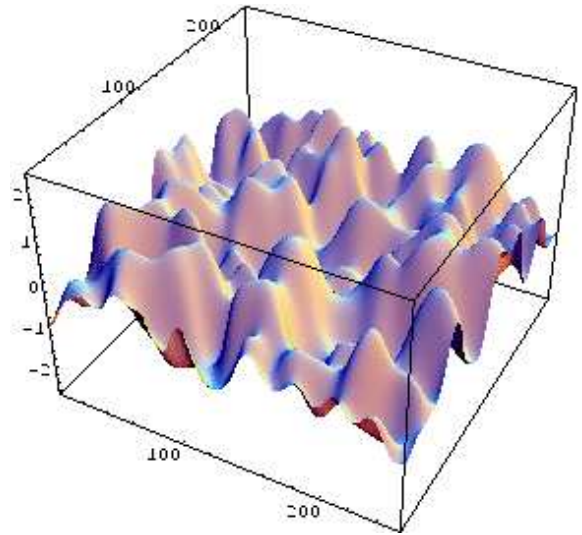}}

\

\caption[Fig001]{\label{onefieldslice} {Growth of quantum fluctuations in the process of symmetry breaking in the quadratic model (\ref{aB1}).  }}
\end{figure}

The growth of fluctuations in this model is shown in Fig.
\ref{onefieldslice}. It shows how fluctuations grow in a
two-dimensional slice of 3D space. Maxima correspond to domains
with $\phi >0$, minima correspond to domains with $\phi<0$.

The dynamics of spontaneous symmetry breaking in this model is even better
illustrated by the computer generated movie is included in this submission as a file 1.gif, which can be found also  at 
http://physics.stanford.edu/gfelder/hybrid/1.gif. It consists of an
animated sequence of images similar to the one shown in Fig.
 \ref{onefieldslice}. These images show the whole process of
spontaneous symmetry breaking from the growth of small gaussian
fluctuations of the field $\phi$ to the creation of domains with
$\phi = \pm v$.

\begin{figure}[Fig001]
\centering \leavevmode\epsfysize=7.5cm \epsfbox{\picdir{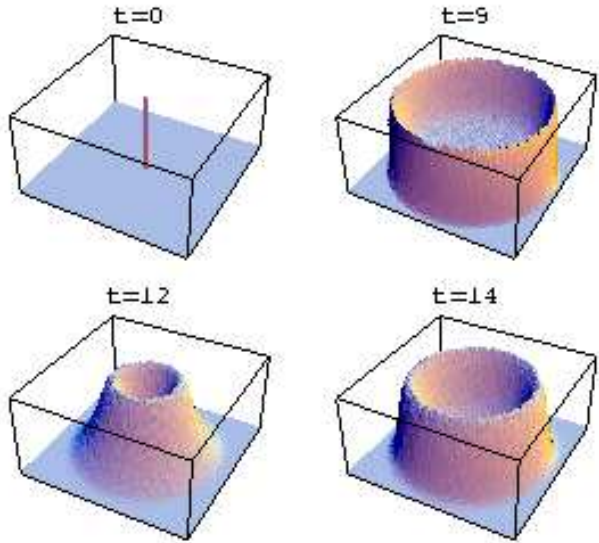}}

\

\caption[Fig001]{\label{complex} {The process of symmetry breaking in the model (\ref{aB1}) for a complex field $\phi$. The field distribution falls down to the minimum of the effective potential at $|\phi| = v$ and experiences only small oscillations with rapidly decreasing amplitude $|\Delta\phi| \ll v$.}}
\end{figure}

Similar results are valid for the theory of a multi-component
scalar field $\phi_i$ with the potential (\ref{aB1}). For example,
the behavior of the probability distribution $P(\phi_1,\phi_2,t)$
in the theory of a complex scalar field $\phi = (\phi_1 +
i\phi_2)/\sqrt 2$ is shown in Fig. \ref{complex}. As we see, after a single oscillation this probability
distribution has stabilized at $|\phi| \sim v$. A
computer generated movie illustrating this process is included in this submission as a file 2.gif, which can be found also  at http://physics.stanford.edu/gfelder/hybrid/2.gif.

\subsection{Cubic potential}

Another important example of tachyonic preheating is provided by the
theory
\begin{equation}
V= -{\lambda\over 3} v\phi^3+{\lambda\over 4}\phi^4+{\lambda\over 12} v^4 \ .
\label{cub}
\end{equation}
This potential is a prototype of the potential that appears in
descriptions of symmetry breaking in F-term hybrid inflation
\cite{Stewart,hybrid2}.

The first question to address concerns the initial amplitude of
the tachyonic modes in this model. This is nontrivial because
$m^2(\phi) = -2\lambda v \phi +3\lambda\phi^2$ vanishes at $\phi =
0$. However, eq. (\ref{aBB}) implies that scalar field
fluctuations with momentum $\sim k$  have initial amplitude
$\langle \delta\phi^2 \rangle \sim {k^2\over 8\pi^2}$. They enter
a self-sustained tachyonic regime if $k^2 < |m^2_{\rm eff}| =
2\lambda v\sqrt{\langle \delta\phi^2\rangle } \sim {\lambda v k
\over 2\pi} $, i.e. if $k < {\lambda v \over 2\pi}$. The average
initial amplitude of the growing tachyonic fluctuations with
momenta smaller than ${\lambda v \over 2\pi}$ is
\begin{equation}\label{typical}
 \delta\phi_{\rm rms} \sim {\lambda v   \over 4\pi^2}.
\end{equation}
These fluctuations grow until the amplitude of $\delta\phi$
becomes comparable to $2v/3$, and the effective tachyonic mass
vanishes. At that moment the field can be represented as a
collection of waves with dispersion $\sqrt{\langle
\delta\phi^2\rangle} \sim v$, corresponding to coherent states of
scalar particles with occupation numbers $n_k \sim
\left({4\pi^2\over \lambda}\right)^2 \gg 1.$

\begin{figure}[Fig001]
\centering \leavevmode\epsfysize=7cm \epsfbox{\picdir{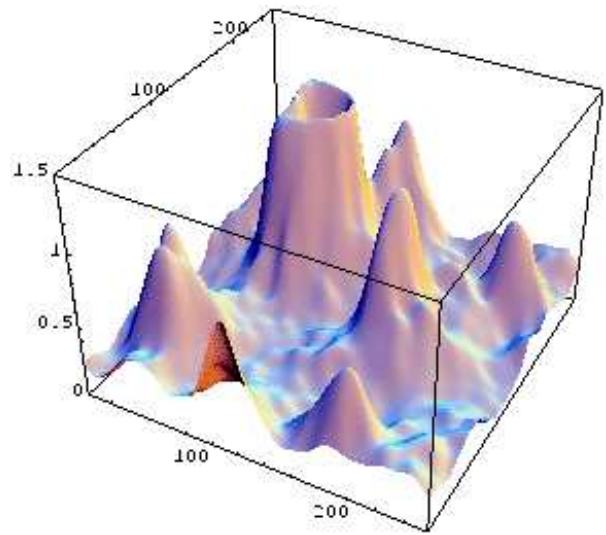}}

\

\caption[Fig001]{\label{2gif} Fast growth of the peaks of the distribution of the field $\phi$ in the cubic model  (\ref{cub}). It should be compared with Fig. \ref{onefieldslice} for the quadratic model (\ref{aB1}).}
\end{figure}

Because of the nonlinear dependence of the tachyonic mass on
$\phi$, a detailed description of this process is more involved
than in the theory (\ref{aB1}). Indeed, even though the typical
amplitude of the growing fluctuations is given by (\ref{typical}),
the speed of the growth of the fluctuations increases considerably
if the initial amplitude is somewhat bigger than (\ref{typical}).
Thus even though the fluctuations with amplitude a few times
greater than (\ref{typical}) are exponentially suppressed, they
grow faster and may therefore have greater impact on the process
than the fluctuations with  amplitude (\ref{typical}).  Low
probability fluctuations with $ \delta \phi \gg \delta\phi_{\rm
rms}$  correspond to peaks of the initial Gaussian distribution of
the fluctuations of the field $\phi$. Such peaks tend to be
spherically symmetric \cite{BBKS}.   As a result, the whole
process looks not like a uniform growth of all modes, but more
like  bubble production (even though there are no instantons in
this model). The results of our lattice simulations for this model
are shown in  Fig. \ref{2gif}. The bubbles (high peaks of the
field distribution) grow, change shape, and interact with each
other, rapidly  dissipating the vacuum energy $V(0)$. A
computer generated movie illustrating this process is included in this submission as a file 3.gif. It can be found also  at http://physics.stanford.edu/gfelder/hybrid/3.gif.

\begin{figure}[Fig002]
\centering \leavevmode\epsfysize=8cm \epsfbox{\picdir{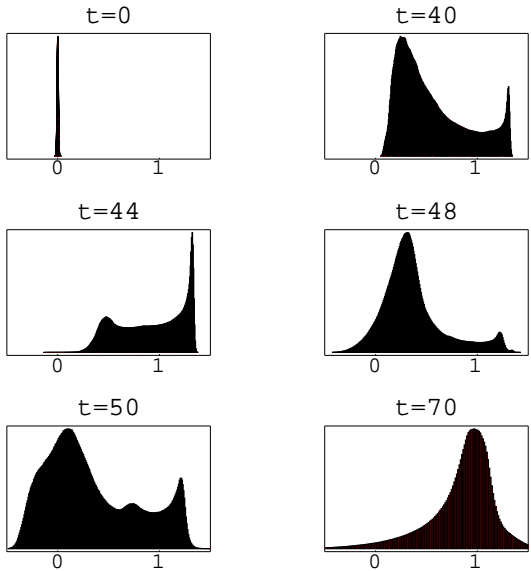}}

\

\caption[Fig002]{\label{cubicdistrib} Histograms describing the process of symmetry breaking in the model (\ref{cub}) for $\lambda = 10^{-2}$. After a single oscillation the distribution acquires the form shown in the last frame and after that it practically does not oscillate.}
\end{figure}

Fig. \ref{cubicdistrib} shows the probability distribution
$P(\phi,t)$ in the model (\ref{cub}). As we see, in this model the
field also relaxes near the minimum of the effective potential
after a single oscillation.

One should note that numerical investigation of this model
involved specific complications due to the necessity of performing
renormalization. Lattice simulations involve the study of modes
with large  momenta that are limited by the inverse lattice
spacing. These modes give an additional contribution to the
effective parameters of the model. In the simple model (\ref{aB1})
these corrections were relatively small, but in the cubic model
they induce an additional linear term $\lambda v \phi
\langle\phi^2\rangle$. This term should be subtracted by the
proper renormalization procedure, which brings the effective
potential back to its form (\ref{cub}).

For completeness we would like to mention also that in the theory
with the quartic potential $V=V(0)-{ 1 \over 4} \lambda \phi^4$
the decay of the symmetric phase occurs via tunneling and the
formation of bubbles, even though there is no barrier between
$\phi = 0$ and $\phi \not = 0 $ \cite{Linde1}.  In this case
quantum tunneling can be heuristically interpreted as a  building
up of stochastic fluctuations $\delta \phi$ \cite{Linde2}.  In
this respect the character of tachyonic instability for the cubic
potential is intermediate between the quadratic and quartic
potentials.

\section{Tachyonic preheating in hybrid inflation}

The results obtained in the previous section have important
implications for the theory of reheating in the hybrid inflation
scenario.  The basic form of the effective potential in this
scenario is \cite{Hybrid}
\begin{equation}\label{hyb_eqn}
V(\phi,\sigma) = {\lambda \over 4} (\sigma^2 - v^2)^2 + {g^2 \over 2}
\phi^2 \sigma^2 + {1 \over 2} m^2 \phi^2 \, .
\end{equation}
The point where $\phi=\phi_c = M/g$   and $\sigma=0$ is a
bifurcation point. Here  $M \equiv \sqrt{\lambda} v$. The
global minimum is located at $\phi=0$ and $|\sigma|= v$. However,
for $\phi > \phi_c$ the squares of the effective masses of both
fields $m_\sigma^2 = g^2 \phi^2 - \lambda v^2 + 3 \lambda
\sigma^2$ and $m^2_{\phi}=m^2+g^2\sigma^2$ are positive and the
potential has a valley at $\sigma=0$.  Inflation in this model
occurs while the $\phi$ field rolls slowly in this valley towards
the bifurcation point.  When $\phi$ reaches $\phi_c$, inflation
ends and the fields rapidly roll towards the global minimum at
$\phi = 0$, $|\sigma| = v$. If $\sigma$ is a real one-component
scalar, this may lead to the formation of domain walls. To avoid
this problem, we assume that $\sigma$ is a complex field. In this
case symmetry breaking after inflation produces cosmic strings
instead of domain walls \cite{Hybrid}.

In realistic versions of this model the mass $m$ is extremely
small,  as well as its initial velocity  $\dot{\phi}$.  The fields fall
down along a certain trajectory $\phi(t), \sigma(t)$ in such a way
that initially this trajectory is absolutely flat, then it rapidly
falls down, and then it becomes flat again near the minimum of
$V(\phi,\sigma)$. This implies that the curvature of the effective
potential along this curve is initially negative. Therefore the
fields should experience tachyonic instability along the way.

The decay of the homogeneous inflaton field and preheating in
hybrid inflation were considered in two papers: for the simplest
non-supersymmetric scenario with a variety of parameters
\cite{hybrid1} and for a SUSY F-term model \cite{hybrid2}. Both
papers were focused on the possibility of parametric resonance.
However,  in \cite{hybrid1} it was also pointed out that for $g^2
\gg \lambda$ the field $\sigma$ falls down only when the field
$\phi$ reaches some point  $\phi \ll \phi_c$. As a result, the
motion of the field  $\sigma$ occurs just like the motion of the
field $\phi$ in the theory (\ref{aB1}). In this case one has a
tachyonic instability and the fields relax near the minimum of
$V(\phi,\sigma)$ within a single oscillation \cite{hybrid1}. For
all other relations between $g^2$ and $\lambda$ the fields follow
more complicated trajectories. One might expect that the fields
would in general experience many oscillations, which might or
might not lead to parametric resonance \cite{hybrid1,hybrid2}.

We performed an investigation of preheating in hybrid inflation in
the model (\ref{hyb_eqn}) with two scalar fields (one real and
one complex) and in SUSY-motivated F-term and D-term inflation
models with three complex fields. We used methods similar to those
that we applied  in the previous section to the investigation of
spontaneous symmetry breaking, including 3D lattice simulations.
We have found that efficient tachyonic preheating is a generic
feature of the hybrid inflation scenario, which means that the
stage of oscillations of the quasi-homogeneous components of the
scalar fields driving inflation is typically terminated by the
backreaction of fluctuations. Here we report the qualitative
results of our findings, leaving the technical details for the
longer paper \cite{GBFKLT}.

\begin{figure}[Fig001]
\centering \leavevmode\epsfysize=7.5cm \epsfbox{\picdir{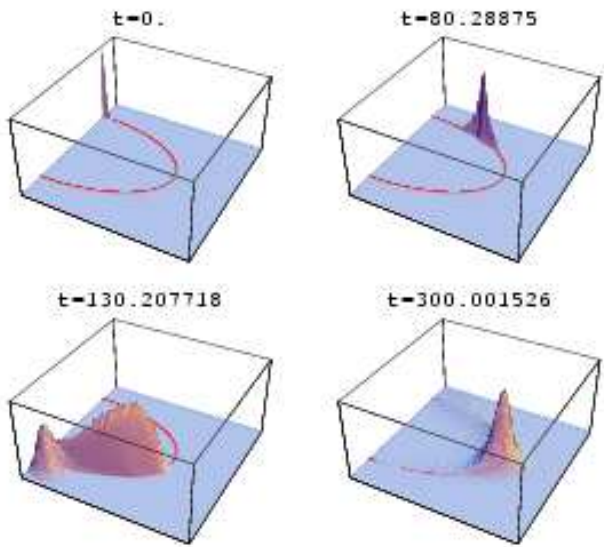}}

\

\caption[Fig001]{\label{basic} The process of symmetry
breaking in the hybrid inflation model (\ref{hyb_eqn}) for $g^2 \ll \lambda$. The field distribution moves along the ellipse $g^2\phi^2 + \lambda\sigma^2 = g^2\phi_c^2$ from the bifurcation point  $\phi = \phi_c$, $\sigma = 0$.}
\end{figure}

In particular, Fig. \ref{basic} shows the process of spontaneous
symmetry breaking in the theory (\ref{hyb_eqn}) for $g^2 = 10^{-4}$,
$\lambda = 10^{-2}$, $M = 10^{15}$ GeV. The probability distribution oscillates along the ellipse  $g^2\phi^2 + \lambda\sigma^2 = g^2\phi_c^2$. As before, it relaxes near
the minimum of the effective potential within a single oscillation. A
computer generated movie illustrating this process is included in this submission as a file 4.gif. It can be found also at http://physics.stanford.edu/gfelder/hybrid/4.gif.

The theory of preheating in D-term inflation for various relations
between $g^2$ and $\lambda$ \cite{Dvali} is very similar to the
theory discussed above. Meanwhile, in the case $g^2 = 2\lambda$
the effective potential (\ref{hyb_eqn}) has the same features as
the effective potential of SUSY-inspired F-term inflation
\cite{Stewart}. In this scenario the fields $\phi$ and $\sigma$
fall down along a simple linear trajectory \cite{hybrid2}, so that
instead of following each of these fields one may consider a
linear combination of them and find the effective potential in
this direction. This effective potential has exactly the same
shape as our cubic potential (\ref{cub}). Thus all of the results
which we obtained for tachyonic preheating in the theory
(\ref{cub}) should be valid for F-term inflation as well, with
minor modifications due to the presence of additional degrees of
freedom that can be excited during preheating. Indeed, we were
able to confirm these conclusions by lattice simulations of the
F-term and D-term models.

Thus we see that tachyonic preheating is a typical feature of
hybrid inflation. The production of bosons in this regime is
nonperturbative, very fast, and efficient, but it is usually not
related to parametric resonance. Instead it is related to the
production and scattering of classical waves of the scalar fields.
Of course, one should keep in mind
that there may exist some particular versions of hybrid inflation
in which tachyonic preheating is inefficient, e.g. because of fast
motion of the field $\phi$ near the bifurcation point.

So far we have discussed tachyonic preheating in the inflaton
sector of hybrid models, which leads to the decay of the
homogeneous fields and the excitation of their fluctuations.
Preheating in the non-inflaton sector and the subsequent
development of equilibrium in hybrid models were considered in
\cite{FK}. Light bosonic fields interacting with scalars from the
inflaton sector are dragged into the process of preheating.
Excitations of these fields rapidly acquire large occupation
numbers and further evolve into equilibrium together with the
scalars from the inflaton sector.

The tachyonic nature of preheating in hybrid inflation implies, in
particular, that instead of the production of gravitinos by a
coherently oscillating field \cite{gravitino1,gravitino2}, in
hybrid inflation one should study gravitino production due to the
scattering of classical waves of the scalar fields produced by
tachyonic preheating.  Our results may also be important for the
theory of the generation of the baryon asymmetry of the universe
at the electroweak  scale~\cite{EWBAR}.

From a more general point of view, however, the most important
application of our results is to the general theory of spontaneous
symmetry breaking. This theory constitutes the basis of all models
of weak, strong and electromagnetic interactions. The new methods
developed during the last few years in application to the theory
of reheating after inflation have been applied in this paper to
the theory of spontaneous symmetry breaking. These methods have
for the first time allowed us not only to calculate correlation
functions and spectra of produced particles, but to actually {\it
see}\, the process of spontaneous symmetry breaking and to reveal
some of its rather unexpected features. We will return to the
discussion of this issue in the coming publication \cite{GBFKLT}.

We thank NATO Linkage Grant 975389 for support. J.G.B.~and
G.F.~thank CITA for its hospitality.  L.K.~and P.G.~were supported
by NSERC; L.K.~was supported by CIAR; G.F. and A.L. were supported
by NSF grant PHY-9870115 and by the Templeton Foundation;
J.G.B.~was supported by Spanish CICYT grant AEN97-1678, and also
thanks the TH-Division at CERN for its hospitality.

\end{document}